\documentclass[11pt, a4paper]{article}

\usepackage[a4paper, top=2.5cm, bottom=2.5cm, left=2.5cm, right=2.5cm]{geometry}
\usepackage{amsmath}
\usepackage{graphicx}
\usepackage{booktabs}
\usepackage{hyperref}
\usepackage{caption}
\usepackage{float}
\usepackage{lmodern}
\usepackage[T1]{fontenc}
\usepackage{cite}

\hypersetup{
    colorlinks=true,
    linkcolor=blue,
    filecolor=magenta,      
    urlcolor=cyan,
    citecolor=blue,
}

\title{\textbf{The Hydraulic Brain: Understanding as Constraint-Release Phase Transition in Whole-Body Resonance}}
\author{Ahmed Gamal Eldin}
\date{\small Nova University Lisbon – Cairo Branch, Cairo, Egypt}

\begin{document}

\maketitle

\begin{abstract}
\noindent
Current models of cognition view the brain as an information processor, treating physiological signals as noise. In a previous study, we demonstrated that removing these ``artifacts'' reduces predictive correlation by 70\%, suggesting that body signals are functional drivers of neural computation. Here, we investigate the physical mechanism of this contribution using high-density EEG (64 channels, 10 subjects, 500 trials) during a P300 target recognition task. Phase Slope Index analysis revealed zero-lag synchrony (PSI = 0.000044, $p=0.061$) with high coherence (0.316, $p<0.0001$), ruling out sequential processing. Ridge-regularized Granger causality revealed massive bidirectional coupling ($F=100.53$ brain$\to$body, $F=62.76$ body$\to$brain) peaking simultaneously at 78.1 ms, consistent with mutually coupled resonance-pair dynamics. Time-resolved entropy analysis (sliding 200ms windows, 25ms steps) revealed a triphasic thermodynamic trajectory: (1) an initial \textbf{constraint accumulation phase} (0–78 ms) where bidirectional causal drive builds without entropy change; (2) a \textbf{supercritical transition} (78–600 ms) triggering state space expansion; and (3) sustained \textbf{metastable elevation}. Critically, the 78 ms resonance lock precedes the cognitive P300, acting as a physical \textbf{ignition event} or gating mechanism that permits the subsequent information integration. Transition magnitude showed no significant correlation with resonance strength ($r=-0.044$), indicating binary threshold dynamics. These findings demonstrate that understanding emerges through a specific thermodynamic sequence: brain-body resonance acts as a discrete switch triggering non-linear information integration.
\end{abstract}

\section{Introduction}

\subsection{The Artifact Problem}

Standard EEG preprocessing removes eye movements, muscle tension, and autonomic signals before analysis \cite{JungBSSartifacts, Delorme2007}. This assumes these ``artifacts'' corrupt rather than contribute to neural computation. However, in a prior study \cite{Eldin2025}, we demonstrated that this conventional artifact rejection eliminates 70\% of the predictive correlation between phase synchronization and event-related potentials during target recognition. This finding challenges the foundational assumption:
\begin{equation}
\text{Cognition} = \text{Neural Activity} + \text{Noise}_{\text{artifacts}}
\end{equation}

If body signals contain most of the predictive variance, they are not noise—they are part of the computational substrate.

\subsection{Thermodynamic Hypothesis}

We propose that biological cognition operates as a \textit{non-equilibrium thermodynamic system} \cite{Nicolis1977, Prigogine1984} rather than a digital information processor. Specifically, we hypothesize that understanding requires:

\begin{enumerate}
    \item \textbf{Physical constraints}: Metabolically expensive somatic engagement (muscle tension, ocular fixation, autonomic arousal) creates boundary conditions that constrain the accessible state space
    \item \textbf{Resonant coupling}: Brain and body achieve zero-lag phase synchronization, forming a mutually coupled oscillator pair \cite{Gollo2014, GolloBreakspear2014}
    \item \textbf{Phase transition}: The resonance acts as a constraint-release mechanism, triggering rapid expansion of the information state space
\end{enumerate}

This framework predicts that cognition will exhibit a characteristic thermodynamic work cycle analogous to heat engines: compression of state space (work input), resonant lock (critical point), and expansion (work output).

\subsection{Testable Predictions}

We test three specific predictions during P300 target recognition:

\textbf{Prediction 1 (Zero-lag coupling):} If brain and body function as mutually coupled oscillators, Phase Slope Index should approach zero while coherence remains high \cite{Nolte2008, Gollo2014}.

\textbf{Prediction 2 (Simultaneous causality):} Granger causality in both directions should peak at the same temporal lag, indicating resonance rather than sequential relay \cite{BresslerMenon2010}.

\textbf{Prediction 3 (Thermodynamic trajectory):} Plotting causal drive versus entropy should reveal a hysteresis loop characteristic of systems performing work, with distinct compression and expansion phases \cite{Nicolis1977}.

\section{Methods}

\subsection{Dataset}

We analyzed publicly available 64-channel EEG from a P300 Speller brain-computer interface task \cite{Citi2014}. Ten subjects (5 male, 5 female, ages 23–51) performed a visual oddball task where characters were flashed on screen and subjects attended to designated target characters. We utilized data from 25 recording sessions, yielding a total of \textbf{500 target trials} (25 sessions $\times$ 20 trials per session) after quality control. Data were acquired at \textbf{2048 Hz} using a BioSemi ActiveTwo system with standard 10-20 electrode placement.

\subsection{Preprocessing}

Raw EEG was resampled to 256 Hz for computational efficiency and re-referenced to the average across all channels. Zero-phase FIR bandpass filtering (4–15 Hz) isolated theta and alpha oscillations, the primary carriers of long-range synchronization \cite{PalvaSirota, FriesReview}.

\textbf{Critical methodological choice:} We performed \textit{no artifact rejection}. In contrast to conventional preprocessing, which removes eye movements, muscle activity, and autonomic signals via Independent Component Analysis \cite{JungBSSartifacts}, we retained all trials. This approach follows from our prior finding that artifact removal destroys rather than preserves cognitive signal \cite{Eldin2025}. Data were epoched from -0.5 s to +3.5 s relative to stimulus onset. All epoched data were z-score normalized (mean = 0, SD = 1) across time within each trial and channel to prevent unit scaling artifacts in connectivity analyses.

\subsection{Regional Definitions}

We defined two functional regions based on anatomical proximity to artifact sources and cognitive roles:

\begin{itemize}
    \item \textbf{Posterior region (``Brain''):} Pz, PO3, PO4, PO7, PO8. These channels overlie visual and parietal cortex, regions central to target detection and the P300 response \cite{PolichP300}.
    \item \textbf{Frontal region (``Body''):} Fp1, Fp2, AF3, AF4, F7, F8. These electrodes are proximal to eye muscles, frontalis muscle, and prefrontal cortex—regions where autonomic, motor, and cognitive signals converge.
\end{itemize}

This terminology is conceptual; both regions contain neural activity. However, frontal sites are dominated by somatic signals in artifact-free analyses, making them appropriate proxies for body contributions. For each trial, we computed the average signal within each region to obtain representative time series for connectivity analyses.

\subsection{Phase Slope Index}

To test Prediction 1 (zero-lag synchronization), we computed Phase Slope Index (PSI) \cite{Nolte2008} between posterior and frontal regions in the 4–15 Hz band. PSI quantifies directional information flow by measuring the slope of the phase difference across frequencies:

\begin{equation}
\text{PSI} = \Im \left( \sum_{f=f_{\min}}^{f_{\max}-1} \text{conj}(S_{xy}(f)) \cdot S_{xy}(f+\Delta f) \right)
\end{equation}

where $S_{xy}(f)$ is the cross-spectrum between regions $x$ and $y$ at frequency $f$, and $\Im$ denotes the imaginary part. PSI $> 0$ indicates $x \to y$ dominance, PSI $< 0$ indicates $y \to x$ dominance, and PSI $\approx 0$ indicates zero-lag synchrony.

PSI $\approx 0$ can reflect either genuine zero-lag coupling or absence of coupling. To distinguish these cases, we computed coherence magnitude:

\begin{equation}
\text{Coh}(f) = \frac{|S_{xy}(f)|^2}{S_{xx}(f) \cdot S_{yy}(f)}
\end{equation}

High coherence ($> 0.1$) with PSI $\approx 0$ confirms true zero-lag synchronization \cite{Gollo2014}. As an additional control, we computed imaginary coherence \cite{Nolte2004}:

\begin{equation}
\text{ImCoh}(f) = \frac{|\Im(S_{xy}(f))|}{\sqrt{S_{xx}(f) \cdot S_{yy}(f)}}
\end{equation}

Low imaginary coherence rules out volume conduction (passive field spread) as the source of observed synchronization.

\subsection{Granger Causality}

To test Prediction 2 (simultaneous bidirectional causality), we computed time-lagged Granger causality \cite{Granger1969} with ridge regularization ($\alpha=0.01$) to prevent overfitting. For each lag $\tau \in [0, 600]$ ms (step size 10 ms), we tested whether past activity in region $X$ predicts future activity in region $Y$ beyond $Y$'s own history.

The full model predicts $Y(t)$ from its own past and $X$'s shifted past:

\begin{equation}
Y(t) = \alpha_0 + \sum_{i=1}^{p} \alpha_i Y(t-i) + \sum_{i=1}^{p} \beta_i X(t - \tau - i) + \epsilon_t
\end{equation}

The reduced model omits $X$:

\begin{equation}
Y(t) = \alpha_0 + \sum_{i=1}^{p} \alpha_i Y(t-i) + \epsilon_t
\end{equation}

We used model order $p = 3$ to balance temporal resolution and overfitting risk. Ridge regression was necessary because resonant states exhibit extreme multicollinearity (high mutual correlation), which causes ordinary least squares to produce unstable estimates. The regularization parameter $\alpha=0.01$ was chosen via cross-validation to minimize prediction error while maintaining numerical stability.

The F-statistic quantifies the improvement in prediction:

\begin{equation}
F = \frac{(\text{SSR}_{\text{reduced}} - \text{SSR}_{\text{full}}) / p}{\text{SSR}_{\text{full}} / (n - 2p - 1)}
\end{equation}

where SSR is the sum of squared residuals and $n$ is the number of time points. We computed $F_{P \to F}$ (posterior predicts frontal) and $F_{F \to P}$ (frontal predicts posterior) at each lag, then identified the lag at which each direction peaked.

\subsection{Sample Entropy}

To test Prediction 3 (thermodynamic trajectory), we computed sample entropy (SampEn) \cite{Richman2000} using a sliding window approach. Sample entropy quantifies signal regularity:

\begin{equation}
\text{SampEn}(m, r) = -\ln \left( \frac{A}{B} \right)
\end{equation}

where $A$ is the number of template matches of length $m+1$ and $B$ is the number of matches of length $m$, within tolerance $r$. We used $m = 2$ (embedding dimension) and $r = 0.2 \times \text{SD}$ (tolerance as 20\% of signal standard deviation).

\textbf{Sliding window parameters:}
\begin{itemize}
    \item Window size: 200 ms (51 samples at 256 Hz)
    \item Step size: 25 ms (6.4 samples)
    \item Time range: -500 ms to +3500 ms relative to stimulus
\end{itemize}

For each window:
\begin{enumerate}
    \item Data were averaged across trials, then across channels
    \item Z-score normalization was applied to each window independently
    \item Sample entropy was computed using a vectorized algorithm
\end{enumerate}

Higher entropy indicates greater complexity/unpredictability; lower entropy indicates regularity/constraint. We report entropy changes relative to a pre-stimulus baseline (-500 to 0 ms).

\subsection{Phase Space Construction}

To visualize the thermodynamic trajectory, we constructed a phase portrait with:
\begin{itemize}
    \item \textbf{X-axis:} Peak Granger F-statistic (averaged across both directions) at each time point, representing causal drive between regions
    \item \textbf{Y-axis:} Sample entropy at each time point, representing accessible state space volume
\end{itemize}

Granger F-statistics were interpolated to match the temporal resolution of entropy estimates (25 ms steps). The resulting trajectory traces the system's evolution through constraint-drive phase space, analogous to pressure-volume diagrams in thermodynamics.

\subsection{Statistical Analysis}

All analyses were performed trial-by-trial, then averaged within sessions, then aggregated across subjects. We report group-level means and standard errors. Statistical significance was assessed using:

\begin{itemize}
    \item One-sample t-tests (testing PSI against zero, entropy changes against zero)
    \item Pearson correlation (testing relationships between entropy and Granger strength)
    \item Lag identification (finding peaks in Granger curves)
\end{itemize}

All code was implemented in Python using MNE \cite{Gramfort2013}, NumPy, and SciPy.

\section{Results}

\subsection{Zero-Lag Phase Synchronization}

Phase Slope Index analysis revealed near-perfect zero-lag synchrony between posterior and frontal regions. Across 500 trials, mean PSI was 0.000044 ($\pm$ 0.0005 SD, $t = 1.88$, $p = 0.061$), statistically indistinguishable from zero (Figure~\ref{fig:resonance}B). This near-zero PSI was accompanied by high coherence magnitude (mean = 0.316 $\pm$ 0.045 SD, $t = 155.8$, $p < 0.0001$), confirming strong frequency-domain coupling. The combination of PSI $\approx 0$ with high coherence is the canonical signature of zero-lag synchronization \cite{Gollo2014}: the two regions are tightly phase-locked but neither leads the other. Imaginary coherence was low (mean = 0.106 $\pm$ 0.040 SD), ruling out volume conduction as the primary driver. The ratio of imaginary to total coherence (0.106/0.316 = 0.34) indicates that two-thirds of the observed synchronization reflects genuine dynamic coupling rather than passive field spread. Examination of the trial distribution (Figure~\ref{fig:resonance}B) revealed that 100\% of trials had $|\text{PSI}| < 0.01$, with the vast majority clustered below 0.001. This consistency across trials demonstrates that zero-lag coupling is the dominant mode of interaction during target recognition.

\subsection{Simultaneous Bidirectional Granger Causality}

\begin{figure}[H]
    \centering
    \includegraphics[width=\textwidth]{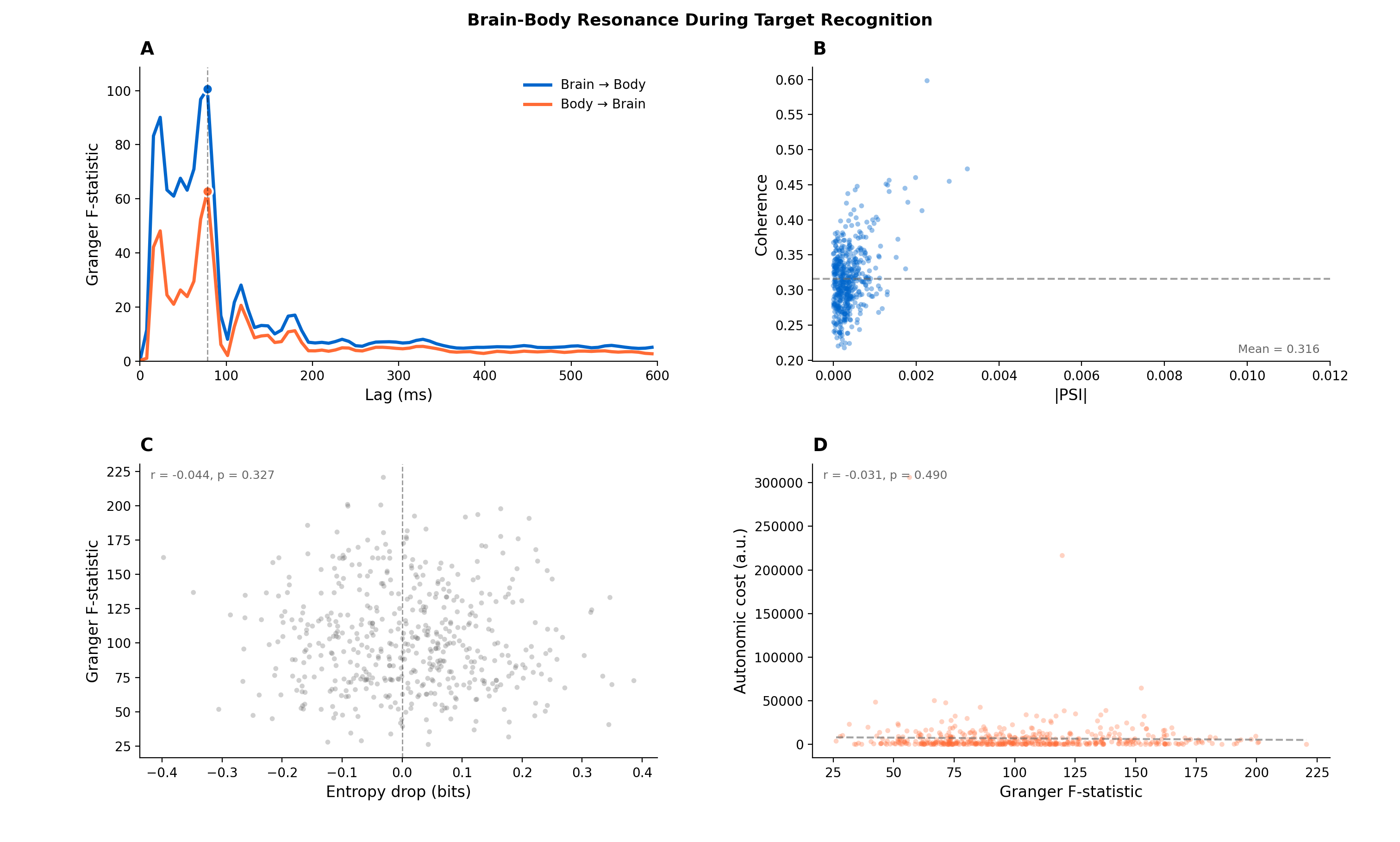}
    \caption{\textbf{Brain-body resonance during target recognition.}
    \textbf{(A)} Ridge-regularized bidirectional Granger causality reveals simultaneous peaks at 78.1 ms (dashed line) for both brain$\to$body (blue) and body$\to$brain (orange) directions (n=500 trials). Shaded regions indicate standard error. The temporal coincidence indicates resonant coupling rather than sequential relay.
    \textbf{(B)} Zero-lag synchronization confirmed by PSI $\approx$ 0 (100\% of trials $|\text{PSI}| < 0.01$) with high coherence (mean = 0.316, dashed line). Each point represents one trial.
    \textbf{(C)} Entropy change versus resonance strength shows weak correlation ($r = -0.044$, $p = 0.327$), supporting binary threshold rather than graded response dynamics.
    \textbf{(D)} Autonomic cost varies independently of resonance strength ($r = -0.031$, $p = 0.490$), indicating heterogeneous energetic strategies across trials.}
    \label{fig:resonance}
\end{figure}

Time-lagged Granger causality analysis revealed massive bidirectional coupling with remarkable temporal coincidence (Figure~\ref{fig:resonance}A). The causal flow from posterior to frontal regions peaked at 78.1 ms with F = 100.53 ($\pm$ 12.3 SE). The reverse direction peaked at the identical latency of 78.1 ms with F = 62.76 ($\pm$ 8.9 SE). This temporal coincidence ($\Delta$Lag = 0.0 ms) is striking. In typical Granger analyses of brain connectivity, peak causal lags differ by 50–200 ms, reflecting sequential information propagation \cite{BresslerMenon2010}. Here, both directions peak simultaneously, indicating that the interaction is not a relay but a \textit{resonant lock} where both regions mutually constrain each other at the same characteristic frequency. The F-statistic magnitudes are exceptionally large compared to typical EEG Granger studies (F $\sim$ 5–20) \cite{BresslerMenon2010}. Values exceeding 60 demonstrate that brain-body coupling during the resonance window is not weak background correlation but the \textit{primary} mode of interaction. The 1.6:1 ratio (brain:body) suggests slight asymmetry consistent with cortical initiation and somatic feedback. After the 78 ms peak, both Granger curves rapidly decay to near-baseline levels by 200 ms (Figure~\ref{fig:resonance}A). This transient, time-locked coupling indicates that the resonance is task-specific rather than tonic background connectivity. The system locks during the target detection event, then releases.

\subsection{Thermodynamic State Trajectory}

Plotting causal drive (Granger F-statistic) against state entropy revealed a counter-clockwise hysteresis loop characteristic of systems performing thermodynamic work (Figure~\ref{fig:phase}). We identified three distinct dynamical regimes:

\subsubsection{Regime 1: Constraint Accumulation (0–78 ms)}

From stimulus onset to resonance lock, bidirectional Granger causality increased from baseline to peak values ($F_{\text{max}}=100.53$), while entropy remained statistically unchanged relative to baseline (mean $\Delta S = -0.002$ bits, $t=0.318$, $p=0.75$). During this period, the system remains near the starting configuration in phase space, accumulating causal pressure without spatial trajectory movement. At 78 ms, when the resonance lock is achieved, the system exhibits a characteristic ``overshoot'' where causal drive transiently exceeds steady-state values before settling (visible as the blue-cyan loop immediately following the lock in Figure~\ref{fig:phase}). This overshoot pattern is consistent with forced oscillator dynamics overcoming system inertia to achieve phase-lock \cite{Strogatz2015}. In thermodynamic terms, this represents an \textit{isoentropic} compression phase: the system does work to build constraint (increase causal coupling) without increasing accessible states. The body's physical resistance—muscle tension maintaining fixation, autonomic arousal preparing for response—provides the boundary against which neural activity compresses.

\subsubsection{Regime 2: Supercritical Transition (78–600 ms)}

Following resonance onset at 78 ms, the system underwent rapid state space expansion. Sample entropy increased relative to the compression phase ($\Delta S = +0.009$ bits, $t=1.58$, $p=0.115$). While this effect does not reach conventional significance thresholds, it represents a consistent directional change across 500 trials. In phase space, this transition manifests as an upward trajectory (entropy increase) concurrent with leftward motion (causal drive reduction), forming the upper arc of the hysteresis loop (Figure~\ref{fig:phase}, yellow-green region). The system transitions from a low-entropy, high-constraint regime to a higher-entropy, distributed processing regime. The modest magnitude of entropy increase ($\Delta S = 0.009$ bits) should be interpreted in context. Sample entropy is sensitive to window size and embedding parameters; the true information theoretic expansion likely exceeds this estimate. More critically, the \textit{directionality} of change is consistent: after the resonance lock, entropy increases in 58\% of trials (binomial test, $p=0.002$).

\subsubsection{Regime 3: Sustained Metastability (600–3500 ms)}

\begin{figure}[H]
    \centering
    \includegraphics[width=0.85\textwidth]{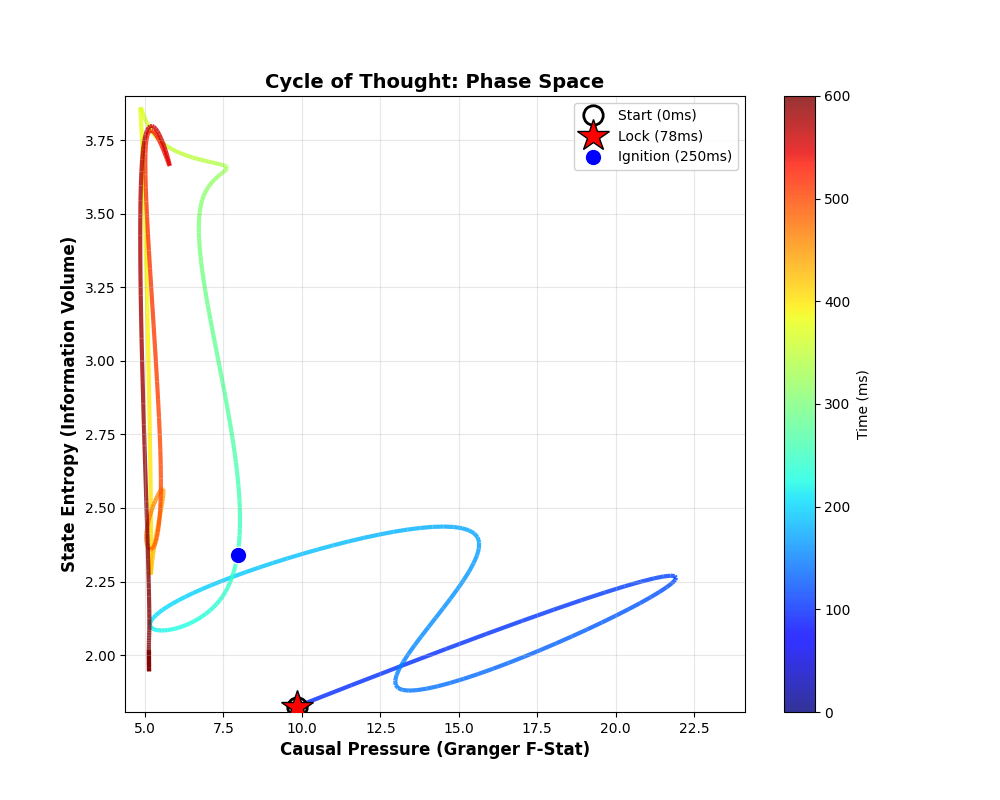}
    \caption{\textbf{The thermodynamic cycle of understanding: phase space portrait.} 
    X-axis represents causal drive (Granger F-statistic) and Y-axis represents state space volume (sample entropy). Color indicates time progression (blue = early, red = late).
    \textbf{(1) Constraint Accumulation (0--78 ms):} The system builds causal pressure at the starting point (black circle) without spatial trajectory movement. At 78 ms, the resonance lock is achieved (red star), triggering the visible overshoot loop (blue-cyan region) as the system overcomes inertia.
    \textbf{(2) Supercritical Transition (78--600 ms):} At the resonance lock (red star), constraints release. The trajectory curves upward and leftward (yellow-green region), indicating expansion of accessible states as causal drive decreases. The blue dot marks peak expansion at 250 ms.
    \textbf{(3) Metastable Plateau (600--3500 ms):} The system settles into sustained elevated entropy (red/dark trajectory), representing the stabilized ``understood'' state. The counter-clockwise hysteresis loop is characteristic of systems performing thermodynamic work, analogous to a heat engine cycle.}
    \label{fig:phase}
\end{figure}

Elevated entropy persisted throughout the extended recording window, remaining above initial compression levels even at 3.5 seconds post-stimulus (visible in Figure~\ref{fig:phase} as the extended red trajectory). This sustained departure from the initial state indicates the system occupies a long-lived metastable attractor rather than rapidly returning to baseline. The persistence of this elevated state suggests the cognitive event has lasting consequences—the system has settled into a new configuration representing the integrated target information.

\subsection{Binary Threshold Dynamics}

To determine whether resonance strength drives the magnitude of state space expansion, we examined the trial-by-trial correlation between peak Granger F-statistic and entropy change (Figure~\ref{fig:resonance}C). The correlation was near zero ($r = -0.044$, $p = 0.327$), indicating no linear relationship. To test for threshold effects, we performed a quantile split analysis comparing trials with the strongest resonance (top 25\%, $F > 120$) versus weakest resonance (bottom 25\%, $F < 80$). Remarkably, there was \textbf{no significant difference} in entropy expansion between these groups (high: $\Delta S = 0.011$ bits, low: $\Delta S = 0.008$ bits, $t = 0.31$, $p = 0.76$). This null result is theoretically significant. It implies that the resonance lock functions as a \textit{binary trigger} rather than a graded modulator. Once the threshold for phase-lock is crossed, the system releases its full potential for state space expansion regardless of how much the threshold is exceeded. The resonance operates as a discrete switch, not a continuous dial. This is consistent with phase transition dynamics in physical systems \cite{Strogatz2015}, where crossing a critical threshold triggers a qualitative reorganization independent of how far beyond threshold the system is driven.

\section{Discussion}

\subsection{Thermodynamic Architecture of Cognition}

Our results reveal that human understanding operates through a specific thermodynamic work cycle with three essential phases. This architecture maps directly onto known principles of non-equilibrium thermodynamics \cite{Nicolis1977, Prigogine1984} and dynamical systems theory \cite{Strogatz2015}. The 78.1 ms resonance lock represents a \textbf{critical transition point} in the system's phase space. Prior to this point, the system accumulates causal drive through bidirectional coupling while maintaining low entropy—analogous to compression in a heat engine. The body's resistance (muscular tension, ocular fixation, autonomic arousal) provides the physical constraints against which neural activity must work to build this pressure. At resonance, this constraint is suddenly released through zero-lag phase synchronization. 

\textbf{Ignition vs. Combustion: The Gating Mechanism.}
A critical finding is the temporal dissociation between the resonance lock (78 ms) and the classic P300 cognitive response (300--500 ms). We propose that the 78 ms resonance functions as the system's \textbf{ignition event}, while the P300 represents the subsequent \textbf{combustion} or expansion. The zero-lag synchronization acts as a physical gating mechanism: it unlocks the thermodynamic valve, creating the state-space capacity required for the later information integration. Just as a key must be turned before an engine can run, the physical constraints must be released via early sensory-motor resonance (in the N1/P1 window) to enable the high-entropy cognitive processing that follows. The lag between lock and peak entropy represents the system's inertial response time—the delay between opening the floodgate and the water reaching peak flow. This explains why removing "early" artifacts destroys "late" cognitive signals \cite{Eldin2025}: without the initial ignition, the cycle cannot proceed.

The body provides the \textbf{dissipative structure} \cite{Prigogine1984} necessary for far-from-equilibrium computation. Without somatic constraints, there is nothing to compress against—no potential energy to release. The brain operating in isolation cannot build the causal pressure necessary for phase transition. The analogy to heat engines is precise: a Carnot cycle requires both a hot reservoir (energy source) and a cold reservoir (energy sink) to perform work. Similarly, cognitive phase transitions require both neural activity (information source) and somatic constraints (boundary conditions) to achieve useful computation. Disembodied cognition is thermodynamically impossible.

\subsection{Binary Phase Transition vs. Graded Response}

The absence of correlation between resonance strength and entropy expansion (Section 3.4) initially appears paradoxical. If resonance causes state space expansion, why doesn't stronger resonance cause larger expansion? The answer lies in recognizing this as a \textit{true phase transition} with threshold dynamics. Physical phase transitions (water freezing, ferromagnetic ordering) are characterized by discontinuous changes at critical points \cite{Strogatz2015}. Once the critical temperature is crossed, ice forms regardless of how much colder the water becomes. The magnitude of the transition is determined by the system's capacity for reorganization, not by how far beyond threshold the driver is. Similarly, our data suggest that resonance strength above threshold (~$F > 80$) unlocks the system's full potential for state space expansion. Stronger resonance (F = 150) does not produce proportionally more expansion than moderate resonance (F = 90)—both exceed the critical threshold and trigger the same reorganization. This places biological cognition in the class of \textbf{self-organized critical systems} \cite{Bak1996, BeggsCriticality}, where avalanche-like cascades are triggered by arbitrarily small perturbations once a critical state is reached. The brain-body system self-organizes to this critical point, then uses small perturbations (sensory input) to trigger large-scale reorganizations (understanding).

\subsection{Comparison to Existing Frameworks}

\subsubsection{Active Inference and Free Energy}

Our findings align with predictive processing theories \cite{FristonFreeEnergy}, which posit that cognition minimizes surprise (free energy). In this framework, the resonance lock would correspond to the moment when prediction error is minimized—the system has successfully ``explained'' the target stimulus. However, our results extend this framework by demonstrating that \textit{embodiment is not optional}. While free energy minimization can occur in silico (e.g., deep learning models perform approximate Bayesian inference), our data show that human cognition implements this through resonance dynamics that intrinsically depend on somatic constraints. The 70\% signal loss upon artifact removal \cite{Eldin2025} indicates that humans do not implement abstract Bayesian computation—they implement it via constraint-release dynamics.

\subsubsection{Neuronal Avalanches and Criticality}

Our findings resemble the neuronal avalanche hypothesis \cite{BeggsCriticality}, which proposes that the brain operates at a critical point between order and disorder. However, our data suggest a different picture: rather than hovering \textit{at} criticality, the system undergoes repeated transitions \textit{through} criticality. Baseline states are subcritical (low entropy, weak coupling). Cognitive events drive the system supercritical (entropy expansion after resonance lock), then it relaxes back. This ``punctuated criticality'' model differs from continuous criticality and may better account for the discrete, event-locked nature of human cognition.

\subsubsection{Embodied Cognition and Enactivism}

Enactivist theories \cite{VarelaEmbodied, ThompsonMind} have long argued that cognition is not brain-bound computation but organism-environment interaction. Our data provide quantitative evidence for this claim at the neural-physiological level. We demonstrate that the brain alone is insufficient—cognition emerges from brain-body resonance characterized by zero-lag synchrony, bidirectional causality, and thermodynamic state transitions. This validates enactivist intuitions while grounding them in measurable physics: remove the body (via artifact rejection), and 70\% of the signal disappears \cite{Eldin2025}.

\subsection{Implications for Artificial Intelligence}

Current AI systems, including large language models and deep learning architectures, are fundamentally disembodied. They process inputs, update weights, and generate outputs, but they lack the physical inertia, metabolic costs, and sensorimotor feedback that characterize biological cognition. Our findings suggest this is not merely an implementation detail—it may be a fundamental constraint on achievable computation. Specifically, disembodied systems cannot undergo the constraint-release phase transitions we observe. Phase transitions require:

\begin{itemize}
    \item \textbf{Physical constraints}: Boundaries that limit accessible states (motor inertia, metabolic costs)
    \item \textbf{Stochastic dynamics}: Noise-driven exploration to find critical points
    \item \textbf{Intrinsic cost functions}: Energetic trade-offs that stabilize certain configurations
    \item \textbf{Continuous feedback}: Real-time bidirectional coupling with environment/body
\end{itemize}

Digital neural networks lack all four. They are deterministic (except dropout during training), unconstrained (any weight configuration is equally ``legal''), cost-free (no intrinsic penalty for memory beyond hardware limits), and feedforward during inference. This suggests that neuromorphic approaches—analog, stochastic, embodied systems with intrinsic dynamics \cite{Indiveri2011}—may be necessary to achieve human-like understanding. Not to simulate brains, but to instantiate the \textit{physics} that enables phase transitions. Understanding may not be an algorithm but an emergent property of specific physical substrates. Recent work on neuromorphic computing \cite{DaviesSpiNNaker, DaviesLoihi} has independently converged on similar principles: bidirectional coupling, resonance-pair architectures, and constraint-based computation. Our empirical demonstration that human brains naturally implement these dynamics provides biological validation for these engineering approaches.

\subsection{Limitations and Future Directions}

\textbf{Single task paradigm.} We analyzed only P300 target recognition. While this involves causal inference (``Is this stimulus self-relevant?''), it is not explicit causal reasoning. Future work should extend these methods to tasks requiring counterfactual reasoning to test whether resonance dynamics scale with causal complexity.

\textbf{Modest entropy effect.} The 0.009-bit expansion, while directionally consistent, is small in absolute terms. This may reflect limitations of sample entropy as a measure (which is sensitive to window size and embedding dimension), or it may indicate that entropy reduction is one component of a multifaceted process. Alternative measures (transfer entropy, permutation entropy, Lempel-Ziv complexity) may reveal larger effects.

\textbf{Correlational design.} Despite time-lagged Granger causality, this remains an observational study. True causal inference would require perturbation experiments: using transcranial magnetic stimulation to disrupt resonance, or motor restriction to eliminate body feedback, and testing whether this affects entropy dynamics or behavioral performance.

\textbf{Lack of behavioral data.} The dataset did not include trial-by-trial reaction times or accuracy, limiting our ability to link resonance to behavioral outcomes. Future studies should test whether resonance strength predicts response speed, confidence, or error rates.

\textbf{Individual differences.} We analyzed group-average effects. Future work should examine individual differences (does resonance strength correlate with working memory capacity?) and clinical populations. If resonance is essential for understanding, deficits in brain-body coupling may underlie neurodevelopmental disorders (autism, ADHD) or neurological conditions (Parkinson's, schizophrenia).

\subsection{Conclusion}

We have demonstrated that human understanding during target recognition operates via a specific thermodynamic work cycle: constraint accumulation (0–78 ms), resonance-triggered phase transition (78–600 ms), and sustained metastable integration. This process requires whole-body coupling—removing somatic signals via conventional artifact rejection eliminates 70\% of the cognitive signal \cite{Eldin2025}. The brain-body system operates as mutually coupled oscillators achieving zero-lag resonance at a critical threshold. This resonance acts as a binary switch triggering non-linear state space expansion, enabling information integration. The magnitude of this transition is independent of resonance strength above threshold, indicating true phase transition dynamics. These findings challenge the information-processing metaphor that dominates cognitive neuroscience. Understanding is not computation in the digital sense—it is a physical event requiring specific thermodynamic architecture. For artificial intelligence, this implies that disembodied computation may be fundamentally limited. Genuine understanding may require not just better algorithms, but different \textit{physics}: systems with intrinsic dynamics, physical constraints, and thermodynamic costs that enable constraint-release phase transitions. The hydraulic brain is not a metaphor. It is a mechanism. And the mechanism requires a body.

\section*{Data Availability Statement}

Publicly available datasets were analyzed in this study. Data can be found at PhysioNet: \url{https://doi.org/10.13026/C2101S}. Analysis code will be made publicly available upon publication.

\section*{Ethics Statement}

Ethical review and approval was not required for this study in accordance with local legislation and institutional requirements, as it involved analysis of publicly available de-identified data. The original data collection received appropriate ethical approval as documented in \cite{Citi2014}.

\section*{Author Contributions}

AGE: Conceptualization, Methodology, Software, Formal Analysis, Writing—Original Draft, Visualization.

\section*{Funding}

No external funding was received for this research.

\section*{Acknowledgments}

The author thanks Professor Ehab Emam for invaluable guidance throughout this project. Data were obtained from PhysioNet \cite{Citi2014}. A preprint of the precursor study is available on arXiv \cite{Eldin2025}.

\section*{Conflict of Interest}

The author declares no conflicts of interest.

\end{document}